%% file: all.tex
\documentclass[conference]{IEEEtran}
\IEEEoverridecommandlockouts
\usepackage{cite} 
\usepackage{listings}

\usepackage{amsmath,amssymb,amsfonts}
\usepackage{algorithmic}
\usepackage{graphicx}
\usepackage{flushend}
\usepackage{textcomp}
\usepackage{xcolor}
\def\BibTeX{{\rm B\kern-.05em{\sc i\kern-.025em b}\kern-.08em
    T\kern-.1667em\lower.7ex\hbox{E}\kern-.125emX}}
\begin{document}

\title{KnowGraph-PM: a Knowledge Graph based Pricing Model for Semiconductors Supply Chains
\\}

\author{\IEEEauthorblockN{Nour Ramzy}
\IEEEauthorblockA{\textit{Leibniz Universität Hannover} \\
Hannover, Germany \\
nour.ramzy@infineon.com}
\and
\IEEEauthorblockN{Sören Auer}
\IEEEauthorblockA{\textit{TIB: Technische Informationsbibliothek } \\
Hannover, Germany \\
soeren.auer@tib.eu}
\and
\IEEEauthorblockN{Javad Chamanara}
\IEEEauthorblockA{\textit{TIB: Technische Informationsbibliothek } \\
Hannover, Germany \\
chamanara@tib.eu}
\and
\IEEEauthorblockN{           }
\IEEEauthorblockA{\textit{        } \\
 \\
}
\and
\IEEEauthorblockN{\centerline{Hans Ehm}}
\IEEEauthorblockA{\textit{Infineon Technologies AG}\\
Munich, Germany \\
hans.ehm@infineon.com
}
}

\maketitle

\begin{abstract}
Semiconductor supply chains are described by significant demand fluctuation that increases as one moves up the supply chain, the so-called bullwhip effect. 
To counteract, semiconductor manufacturers aim to optimize capacity utilization, to deliver with shorter lead times and exploit this to generate revenue. 
Additionally, in a competitive market, firms seek to maintain customer relationships while applying revenue management strategies such as dynamic pricing. Price change potentially generates conflicts with customers. 
In this paper, we present KnowGraph-PM, a knowledge graph-based dynamic pricing model. 
The semantic model uses the potential of faster delivery and shorter lead times to define premium prices, thus entail increased profits based on the customer profile. 
The knowledge graph enables the integration of customer-related information, e.g., customer class and location to customer order data. 
The pricing algorithm is realized as a SPARQL query that relies on customer profile and order behavior to determine the corresponding price premium. 
We evaluate the approach by calculating the revenue generated after applying the pricing algorithm. Based on competency questions that translate to SPARQL queries, we validate the created knowledge graph. 
We demonstrate that semantic data integration enables customer-tailored revenue management. 
\end{abstract}


\section{Introduction}
\input{1-introduction.tex}
\input{2-Background}
\input{3-Implementation}

\input{4-evaluation}

\input{5-conclusion}

\section*{Acknowledgment}

This work has received funding from the EU Electronic Component Systems for European Leadership (ECSEL) Joint Undertaking within Integrated Development 4.0 project (idev40).

\bibliography{bibliography} 
\bibliographystyle{ieeetr}

\end{document}

%% file: 1-introduction.tex
The semiconductor industry is characterized by complex supply chain structures due to its wide range of customers with fluctuating demands for products, long production lead times, short product life cycles and rapidly changing technologies~\cite{mousavi2019simulation}. 
This entails specific phenomena such as the bullwhip effect that increases the variability of orders as they move up the supply chain from retailers to wholesalers to manufacturers to suppliers~\cite{sucky2009bullwhip}. 
Consequently, matching the demand and capacity requires effective planning and optimal utilization of production capacities. 
Moreover, manufacturing lead times are longer than customer order lead times (the time interval between order entry and requested delivery date), thus fulfilling order requests that go beyond promised delivery time is costly~\cite{ott2013granularity}. 
Therefore, to keep utilizing capacity in a profitable way and guarantee customer satisfaction, companies resort to revenue management ideas such as dynamic pricing~\cite{seitz2016robust}, i.e., exploiting faster delivery to generate revenue.
Potentially, conflicts with customers arise, affecting their satisfaction and harming the firm’s long-term relationship with the customer and ultimately its success~\cite{wirtz2003revenue}. 

In fact, to maintain advantage especially needed within the competitive electronics market~\cite{monch2018survey}, semiconductor manufacturers strive to deliver and maintain various customer relationships, and design customized portfolios with distinct values to differentiate between customers' requirements~\cite{chou2014does}. 
Yet, revenue management and customer account management are often conducted independently of each other~\cite{wang2014framework}, although, the flow of information in a supply chain is crucial for carrying out effective exchanges between parties~\cite{stefansson2002business}. 
For instance, centralized information and elaborate information integration among supply chain partners can limit the increase in demand variability i.e., reducing the bullwhip effect~\cite{chen2000quantifying}.

In this article, we propose KnowGraph-PM a pricing solution to measure a price change to each customer for an order. The knowledge graph, depicting customers and orders, subsumes data about customer order behavior as well as customer profile. The former describes order lead times for a customer, e.g., requested, confirmed, while the latter defines customer classes expressing the relationship with the firm.  
For each customer, we calculate a price premium based on lead time variations as well as the corresponding price adjustment factor derived from the respective customer class. For each order placed by a customer, we apply the customer price premium to get the new price for this order. The approach enables, via a knowledge graph, semantic data integration by combining customer demand and order-related data with customer portfolio information. We apply the proposed approach to a use case of an international semiconductor manufacturing firm; the results show that KnowGraph-PM approach allows tailored revenue generation according to customers' profile thus reducing the risk of harming relationships.

The remainder of the paper is organized as follows: 
In Section~\ref{background} we introduce the domain and the key terminology and motivation behind the work. 
We present the approach and the KnowGraph-PM implementation  in Section~\ref{implementation}. 
In Section~\ref{evaluation} we evaluate the implemented solution and discuss the advantages and limitations of our approach.
Finally, we conclude and present an outlook for further steps in Section~\ref{conclusion}. 

%% file: 2-Background.tex
\section{Background and Motivation} \label{background}

\subsection{Background} 
\subsubsection{Lead Time}

\begin{figure}[htbp]
\includegraphics[width=\linewidth]{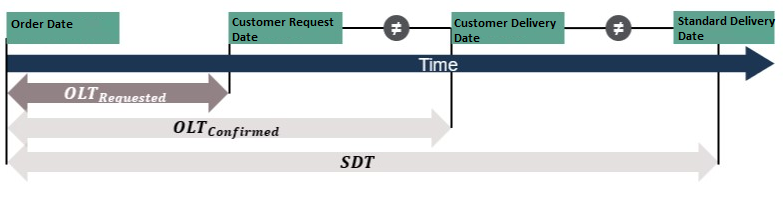} 
\caption{The definitions of lead times.}
\label{fig:lt}
\end{figure}

The \textit{order lead time} (OLT) at each stage of the order, is the time frame between the entry of an order to the time point it passes a designated stage ~\cite{oner2018dynamic} as shown in Figure~\ref{fig:lt}. 
A \textit{requested order lead time} $OLT_{Requested}$ is the difference between \textit{Customer Request Date} for an order, i.e., when the customer wishes to receive the order and the \textit{Order Date} (when the customer placed the order). 
A \textit{confirmed order lead time} $OLT_{Confirmed}$ is the difference between \textit{Customer Delivery Date} i.e., the confirmed order delivery date and the \textit{Order Date}. 
\textit{SDT} is defined as the difference between \textit{Standard Delivery Date} and \textit{Order Date}. It is the lead time the customer can expect till the order is delivered. 
In practice, customers request their orders earlier than \textit{SDT}, i.e., \( SDT > OLT_{Requested}\). 
Moreover, due to the high-competitive market of semiconductors, manufacturers usually sell their products with a shorter $OLT_{Confirmed}$, closer to $OLT_{Requested}$. 
The potential of Lead Time-Based Pricing (LTBP) is seen in the difference between \textit{Standard Delivery Date} and \textit{Customer Delivery Date}. Namely, faster
delivery, $SDT>$ $OLT_{Confirmed}$, is exploited by offering
LTBP as part of dynamic pricing.

\subsubsection{Lead Time-based Pricing (LTBP)}
Namely, dynamic pricing describes the firm’s practice to charge various customers different prices for the same products~\cite{wittman2018customized}. 
For instance,~\cite{seitz2016robust} proposes a supply chain planning framework for revenue management that consists of solutions for demand steering and dynamic pricing where the pricing algorithm relies on the data from order lead time measurement. 
However, LTBP models to identify the price premium based on faster delivery are still under development. 
In fact, some rely on simple mathematical functions, e.g., linear, concave, convex. 
For instance,~\cite{berger2021} introducing Equation~\ref{eq:conv}, explains the rationale behind a convex function; both opportunity costs for the manufacturer and for the customer increase with decreasing lead time.
\begin{equation}
P_{Premium} = \alpha * \log(\frac{OLT_{Confirmed}}{SDT})
\label{eq:conv}
\end{equation}
where $\alpha$ is a factor obtained from simulating customer order behavior for maximizing revenue for the company and $P_{Premium}$ is the price change or the added portion to the original price of the order.
In fact, revenue management (RM) is built upon heterogeneity between customers; frequent buyers need to be treated differently than occasional ones~\cite{hailwood2003customer}. 
Typically, Customer Relationship Management (CRM) enables customer profiling and delimits characteristics for customers and how they can be used to determine the price as part of RM~\cite{zhang2012price}. 

\subsubsection{Customer Relationship Management (CRM)}
 
CRM refers to building one‐to‐one relationships with customers that can drive value for the firm~\cite{kumar2010customer}. 
Consequently, customization of requirements increases maintenance of customers~\cite{johnson2012customer}. 
Namely, by segmenting customers into portfolios, an organization can better understand the relative importance of each customer to the company's total profit~\cite{thakur2016customer}.  
Customer Portfolio Management (CPM) is based on financial performance as well as strategic and economic criteria, e.g., customer account types or classes, regional importance. 
For each criterion, the company chooses measurable characteristics to segment customers accordingly. 
For example, customer classes distinguish customers according to their impact on business, based on the average revenue for the current and previous fiscal year, the volume of purchases, potential sales, the prestige of the account and market leadership.

Especially in semiconductor manufacturing, a competitive domain, meeting customer-specific requirements incites maintaining close associations with customers to identify their specific needs. 
In fact, CPM affects new product development~\cite{yli2008customer}. Especially with the short product life cycle trait for this domain, it accentuates the need for manufacturers to act fast and maintain good relations with the customer. 

\subsection{Motivation}

The bullwhip effect, a significant characteristic of the semiconductor supply chain, entails high inventory capacity, possibly unnecessarily. 
However, to keep utilizing capacity in a profitable way companies exploit revenue management ideas such as dynamic pricing. 
Yet, LTBP strategies are myopic because the decisions are optimized considering solely the profit that is expected to be obtained from the currently arriving customer without any foresight on the long-run impact of these decisions~\cite{oner2018dynamic}. 
Companies realize the need to keep a good relationship with the customer while generating revenues. 

Typically, interactions between revenue management and key account management have been largely ignored as both are often conducted independently of each other~\cite{wang2014framework}. 
In addition,~\cite{li2013controlling} explains how limited information sharing increases the difficulty of reducing the bullwhip effect and leads to inefficient supply chain management. 
Also, information integration increases acquisition and maintenance of customers according to profitability~\cite{johnson2012customer}. 
Moreover, to the best of our knowledge, previously developed LTBP models solely rely on customer order behavior while not including the customer portfolio and specific characteristics.
In this work, we create a knowledge graph that semantically integrates data from different sources, i.e., customers, orders as well as customer account types. 
This solution suggests information integration from various data sources to customer-specific  revenue management via LTBP. 
This improves revenue generation while maintaining customer relationships.

%% file: 3-Implementation.tex
\section{Implementation } \label{implementation}
This section describes the approach we implemented, integrating data from customer relationship management and customer order data. 

\subsection{Knowledge Graph (KG)}\label{AA}
\subsubsection{Domain Ontology}
\begin{figure}[htbp]
\includegraphics[width=\linewidth]{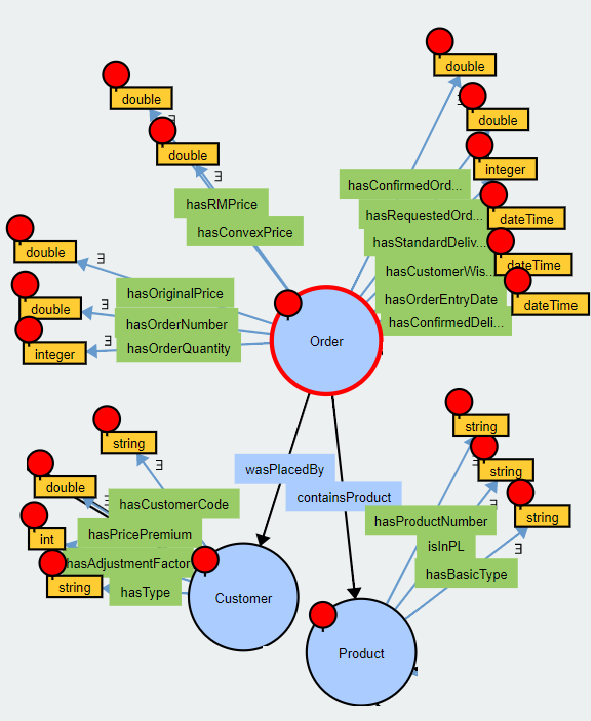} 
\caption{Domain knowledge graph comprising the order, customer and product classes with associated properties.}
\label{fig:KG}
\end{figure}

We use the ontology-based data access (OBDA) approach for semantic data integration. 
The schema is given in terms of an ontology representing the formal and conceptual view of the domain~\cite{de2018using}. 
Figure~\ref{fig:KG} shows the ontology comprising the \textit{Order} class, which enables the coupling of the remaining classes of the domain, e.g., \textit{Product} and \textit{Customer}. 
The \textit{Order} entity is uniquely identified via an \textit{Order Number} and is described via twofold data properties. 
First, order information properties such as the order quantity and the original price of the order. 
Second, an order is described with lead time properties, e.g., order entry date, requested delivery date, and the standard delivery time.
Through its two object properties \textit{containsProduct} and \textit{wasPlacedBy}, an order is linked to the \textit{Product} and \textit{Customer} entities respectively. 
The \textit{Customer} class describes a customer, by assigning each customer a distinct customer code and by categorizing a customer into the defined customer account type or class. 
This categorization into a customer account type is then quantified with the data property \textit{hasAdjustmentFactor}, which assigns each customer a specific pre-defined value based on their respective type. 
Finally, each product is uniquely identified with the help of the product number. 
Also, a \textit{Product} includes data properties that contain all relevant product information such as the product basic type and the product line. 

\subsubsection{Data Sources: Use Case Description}
The use case described here belongs to a international semiconductor manufacturing firm. 
We utilize these two datasets: \textit{DS1} that contains data about order details, e.g., lead times, products and corresponding customers. 
We filter orders containing products within one product line. 
We process roughly 65 thousand orders in the time span of 2016-10-04 to 2020-09-01. 
\textit{DS2} contains customer portfolio, i.e., customer account type, class and location. 
It subsumes 177 customers and corresponding account types (regular, key and others).  
Moreover, we use the corporate memory tool as a platform for data integration\footnote{https://eccenca.com/products/enterprise-knowledge-graph-platform-corporate-memory}. 

\subsection{Lead Time-based Pricing Algorithm}
\begin{figure}[htbp]
\includegraphics[width=\linewidth]{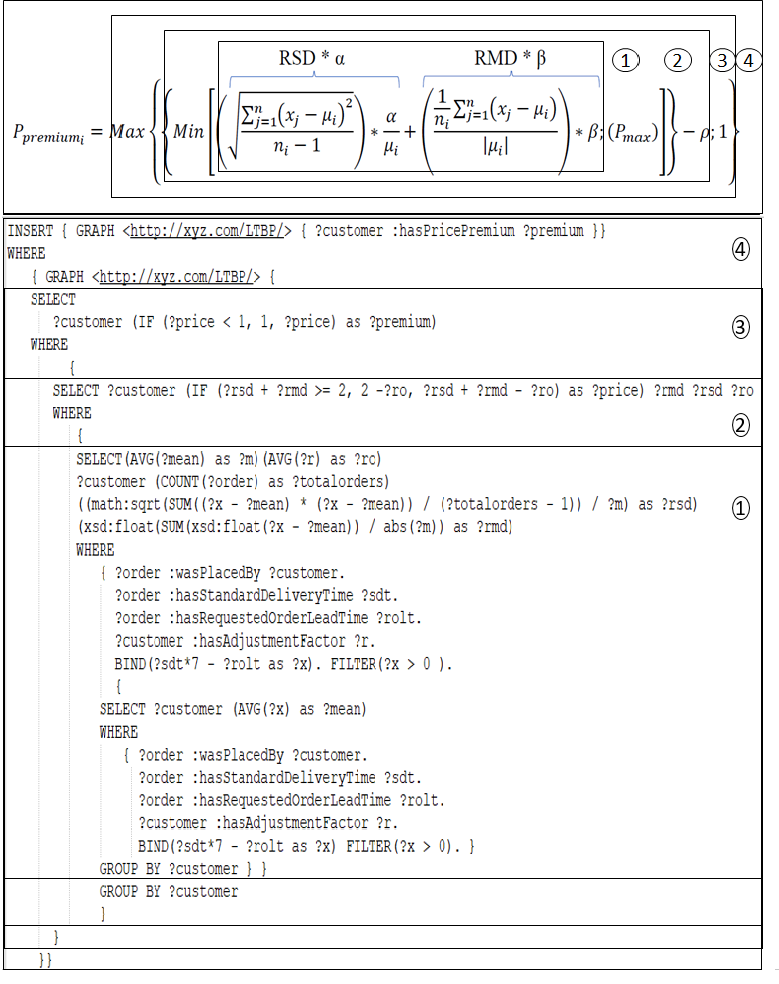} 
\caption{Lead time-based pricing algorithm with the formula (top) and corresponding SPARQL query implementation (bottom).}
\label{fig:algo}
\end{figure}

The pricing algorithm is implemented as a SPARQL query that emulates the equation developed by~\cite{niklas2017}. 
SPARQL\footnote{https://www.w3.org/TR/rdf-sparql-query/} is the query language standardized by the W3C for querying knowledge graphs.
Figure~\ref{fig:algo} describes how parts of the equation match to the nested query shown. 
Steps [1-4] depict how the customer order behavior and customer account type contribute to calculating the price premium \(P_{premium}\) for a specific customer.  
The last step is to calculate the new revenue management price $P_{RM{j}}$ of a specific order \textit{j} knowing the \(P_{premium}\) of this customer. 

\begin{enumerate}
    \item We calculate the relative standard deviation (RSD) and the relative mean deviation (RMD) for each customer individually. This is done only if revenue management is allowed, i.e., \( SDT > OLT_{Requested}\), when customers request their orders earlier than \textit{Standard Delivery Date}. This represents the customer order behavior from previous orders. $\alpha$ and $\beta$ are coefficients to offer the model user the chance to emphasize one parameter more than the other. For instance, one could argue that the average deviation or mean is less important. For simplicity, alpha and beta are chosen to be equal to 1.
    \item The sum of weighed RSD and RMD is then compared to a pre-determined threshold \(P_{max}\) which in this case is set to 2 as this represents the ceiling for the highest premium that can be offered to a customer which is double the original price. The sum is modified by the revenue adjustment parameter \(\rho\). As shown in Table~\ref{tab1}, for each customer account type or class there exists a corresponding \(\rho\) based on revenue ranges. Large revenue ranges, entail more important customer accounts, thus have greater advantage equivalent to higher \(\rho\) leading to smaller premium price.
    \item In order to ensure that no customer receives a premium price smaller than the original we select the max between our first part of the formula and 1.
    \item We assign \(P_{premium}\) to each customer \textit{i} where $ 1<P_{premium}<2$
    \item Upon a new order \textit{j}, we apply Equation~\ref{eq:revmn} to get the revenue management price $P_{RM{j}}$ from the original price $P_{O{j}}$ of the order and customer-specific \(P_{premium}\) assigned in the previous step. As mentioned, the difference between \textit{SDT} and $OLT_{Confirmed}$, is exploited for revenue management.
\end{enumerate}
\begin{equation}
\resizebox{\columnwidth}{!}{ $
P_{RM{j}}= P_{O{j}} + P_{O{j}} [(1- \frac{OLT_{Confirmed_j}}{SDT{j}} )* (P_{premium}-1)]
\label{eq:revmn}$  }
\end{equation}
 
\begin{table}[htbp]
\caption{Exemplary values of price adjustment factor}
\begin{center}
\resizebox{\columnwidth}{!}{
\begin{tabular}{|c|c|c|}
\hline
\textbf{Revenue (r) Range}&\textbf{Customer Class} &\textbf{Price Adjustment Factor \(\rho\) } \\
\hline

 10$<$r$<$100m &Key&0.1 \\
 5$<$r$<$10m&Regular&0.05 \\
 0$<$r$<$5m&Others& 0.025  \\
\hline

\end{tabular}
\label{tab1}
}
\end{center}
\end{table}

%% file: 4-evaluation.tex
\section{Evaluation} \label{evaluation}
The implemented approach combines data from customer relationship management with customer order behavior. 
We use OBDA to represent the domain and create a knowledge graph. 
We evaluate KnowGraph-PM twofold. 
First, we check if the created knowledge graph covers the domain in question. 
Second, we calculate the total revenue generated using static and dynamic pricing  algorithms and compare them with our KnowGraph-PM model.

\subsection{Competency Questions and KG Evaluation}
\begin{figure*}[htbp]
\includegraphics[width=\linewidth]{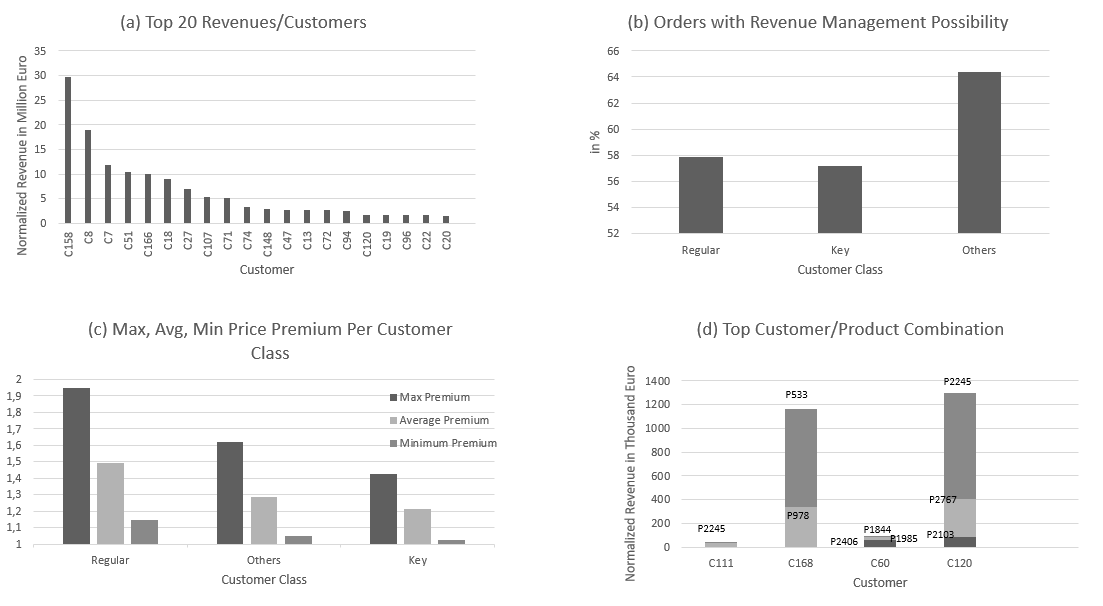} 
\caption{Chart visualization of SPARQL query results.}
\label{fig:graphs}
\end{figure*}

We translate the listed competency questions, defined by domain experts, to SPARQL queries. 
We execute them on the described knowledge graph in Section~\ref{implementation}. 
The results are extracted and represented in Figure~\ref{fig:graphs}.

\textbf{\textit{CQ1: Top 20 most profitable customers:}} 
This competency question retrieves the top 20 customers that generate the highest revenues who are assigned a price premium and who buy the order with the price change. 
The results shown in Figure~\ref{fig:graphs}(a) give a valuable insight into the most influential customers for a firm. 
This can be extended by examining these customer classes and accordingly tailor marketing strategies.

\textbf{\textit{CQ2: Occurrence based customer ranking}:}
 We can examine the overall customer order behavior of the different customer classes. 
 For this, we rank customer types based on the highest percentage of occurrences of \( SDT > OLT_{Requested}\) (RM is possible).
 We analyze, which customer classes are more likely to request a delivery date that is earlier than the \textit{Standard Delivery Date}. 
 This makes them eligible for revenue management with the adjustment factor $\rho$ in Table  ~\ref{tab1} suitable to their respective customer class. 
 The results in Figure~\ref{fig:graphs}(b) show that \textit{Key} account customers are least likely to request an early delivery.
 This is an expected outcome. 
 In fact, key customers have strong relationship established with the company. 
 Thus, they are more likely to enter long-term agreements and contracts entailing stable order behavior i.e., limited variation in lead times.
 Additionally, because of the frequency of their past orders, they can accurately forecast their needs and how a firm can fulfill them.
 Accordingly, this constrains dynamic pricing and revenue management potential.  
            
\textbf{\textit{CQ3: Per customer class price premium estimation:}}
This question provides information about the customer class that is most likely to be the most profitable upon applying revenue management.
The information contains the highest, the lowest and the average Price Premium that will be paid by customers of each customer class.
Also, it gives insights into the customer order behavior of the different customer classes. 
We observe in the results shown in Figure~\ref{fig:graphs}(c) that the Regular customer type has the highest maximum, average, and minimum price premiums. 
This means that this customer class would yield the highest average revenue when the dynamic pricing model is applied. 
       
\textbf{\textit{CQ4: Initial customer and product selection:}}
This CQ determines the combination of customers and products that should be released with the first practical implementation of the LTBP model in order to maximize the initial potential revenue increase using the adjusted prices.
Results in Figure~\ref{fig:graphs}(d) provide the combination of customers and products in terms of profitability and revenue generation in the case where the lead time-based pricing is applied. 

\subsection{Revenue Management (RM)}
Dynamic pricing as part of RM is about generating revenue for the company.
Existing dynamic pricing algorithms, e.g., convex model consider customer behavior while ignoring customer profiles. 
KnowGraph-PM customizes the RM price based on customer order behavior and customer relationship with the company.
To evaluate this, we execute the query in Listing~\ref{lst:label} where we sum the order prices with and without applying revenue management. 
The results show that
\( TotalOrginalPrice < TotalRMPrice < TotalConvexPrice \).
In that, \textit{TotalOrginalPrice} is for the case that no RM is applied, which indeed is the summation of original prices for each order $P_{O{j}}$ as provided in Equation~\ref{eq:revmn},
\textit{TotalRMPrice} is the sum of RM price, i.e., $P_{RM{j}}$ as of Equation~\ref{eq:revmn}, 
and \textit{TotalConvexPrice} is the sum of order prices if Equation ~\ref{eq:conv} is applied.

These results indicate that by applying KnowGraph-PM, a company can generate revenue while taking into consideration the customer portfolio. 
Other models such as the convex generate higher revenue but can entail a disturbed relationship with the customer. 
 The implemented approach integrates data from customer relationship management as well as customer order data to generate a price premium that fits the customer portfolio.
The proposed solution is portable and can be applied to other domains with the right mapping between data and the ontology. 
Additionally, the two evaluation techniques show that it is up to the company to adapt pricing strategies based on customer portfolio. 
SPARQL queries can be tweaked to show details about a specific customer and consequently tailor marketing strategies.
\begin{lstlisting}[caption={Evaluation query},label={lst:label},basicstyle=\small\ttfamily]
SELECT 
  (sum(?rmprice) as ?TotalRMPrice) 
  (sum(?orignalprice) as ?TotalOrginalPrice) 
  (sum(?convexprice) as ?TotalConvexPrice)
FROM <http://xyz.com/LTBP/>
WHERE {
  ?order  :hasRMPrice       ?rmprice. 
  ?order  :hasOriginalPrice ?orignalprice.
  ?order  :hasConvexPrice   ?convexprice. 
}
\end{lstlisting}
 \label{code}

\subsection{Discussion}

The implemented approach integrates data from customer relationship management as well as customer order data to generate a price premium that fits the customer portfolio.
The proposed solution is portable and can be applied to other domains with the right mapping between data and the ontology. 
Additionally, the two evaluation techniques show that it is up to the company to adapt pricing strategies based on customer portfolio. 
SPARQL queries can be tweaked to show details about a specific customer and consequently tailor marketing strategies.  

However, the approach has some limitations. 
First,  we identify a limitation in the evaluation as we compare our model to a convex model with \(\alpha\) chosen to be -0.5 in Equation ~\ref{eq:conv}. 
This choice is not accurate as the parameter is set after using simulation models to optimize the revenue for a specific set of customers. 
Second, for the evaluation, we calculate the total revenue with an assumption that all customers accept the price change. However, in practice, customers can refute the premium price and stick to the original.

%% file: 5-conclusion.tex
\section{Conclusion and Outlook} \label{conclusion}
Semiconductor supply chains are strongly affected by the bullwhip effect, i.e., increasing demand fluctuation.
Thus, in a competitive market, semiconductor manufacturers seek to optimize capacity utilization, deliver with shorter lead times than the agreed contractual times and guarantee customer satisfaction and loyalty. 
Firms exploit faster delivery by resorting to revenue management ideas such as dynamic pricing. 
However, this potentially affects relationships with customers. 

In this work, we propose KnowGraph-PM, a knowledge graph lead time-based pricing approach allowing tailored revenue generation according to customers' profile thus reducing the risk of harming relationships. 
We present dynamic pricing as part of revenue management taking into consideration customer relationship with the customer. Using a pricing algorithm, implemented as a SPARQL query, we assign a price premium to each customer based on their order behavior and changes in the order lead time. Consequently, we calculate a price change for a customer's order. 
We apply the proposed approach to a use case of an international semiconductor manufacturing firm and proved that semantic data integration enables customer profiling aware lead time-based pricing.

As future work, we aim to integrate into KnowGraph-PM an acceptance rate distribution to model customer's behavior towards dynamic pricing. This would lead to more realistic figures in the evaluation thus overcoming the mentioned limitation.
Moreover, we propose extending the knowledge-graph with more contextual data about the customer.
For instance, Customer Master Data (address data, contract data) is an enabler for all customer-driven operations. 
Thus, it can be used to determine the strategic internal decision to treat the customer in a strategic manner, e.g., customer prioritization in times of allocation. 
Consequently, further pricing models that incorporate more factors about a customer can be explored.